\documentclass[journal]{IEEEtran}
\usepackage{ifpdf}
\usepackage{cite}
\ifCLASSINFOpdf
 \usepackage[pdftex]{graphicx}
\else
 \usepackage[dvips]{graphicx}
\fi
\usepackage{amsmath,amssymb,amsfonts}
\usepackage[ruled]{algorithm2e}
\usepackage{algcompatible}
\usepackage[colorinlistoftodos]{todonotes}
\usepackage{tikz}
\usepackage[mathscr]{euscript}
\usepackage{pgfplots}
\usepackage{pgfplotstable}
\usepackage{circuitikz}
\usepackage{graphicx}
\usepackage{textcomp}
\usepackage{float}
\usepackage{array,multirow}
\usepackage{caption}
\usepackage{subcaption}
\usepackage{booktabs}
\usepackage{soul}
\usepackage{diagbox}
\newcommand{\ts}{\textsuperscript}
\allowdisplaybreaks
\newcolumntype{C}[1]{>{\centering\let\newline\\\arraybackslash\hspace{0pt}}m{#1}}
\usetikzlibrary{shapes,arrows,positioning,calc}
\floatstyle{plaintop}
\restylefloat{table}
\def\BibTeX{{\rm B\kern-.05em{\sc i\kern-.025em b}\kern-.08em
    T\kern-.1667em\lower.7ex\hbox{E}\kern-.125emX}}
\pgfmathdeclarefunction{gauss}{2}{\pgfmathparse{1/(#2*sqrt(2*pi))*exp(-((x-#1)^2)/(2*#2^2))}}
\newcolumntype{L}{>{\arraybackslash}m{3cm}}
\begin{document}

%\onecolumn
%\input{coverLetter.tex}

\title{Optimization of DSP Applications Using Parameterized Error Models for Low Power Approximate Adders
}

\author{Celia~Dharmaraj,~\IEEEmembership{Student~Member,~IEEE,}
        Vinita~Vasudevan,~\IEEEmembership{Member,~IEEE,}
        and~Nitin~Chandrachoodan,~\IEEEmembership{Member,~IEEE}% <-this % stops a space
\thanks{The authors are with the Department of Electrical Engineering, Indian Institute of Technology Madras, India.\protect\\
E-mail: {ee13d003,vinita,nitin}@ee.iitm.ac.in.}}

\maketitle

\begin{abstract}
Approximate circuit design has gained significance
in recent years targeting error tolerant applications. In this
paper, we first demonstrate that the 
commonly used assumption that the inputs to the adder are uniformly
distributed results in an inaccurate prediction
of error statistics for multi-level circuits. To overcome this
problem, we derive parameterized error models that can be used within
any optimization framework in order to optimize the number of approximate bits.
We also show that in order to accurately compute the MSE, the optimization framework needs to take into account not just the functionality of
the adder, but also its position in the circuit, functionality of its parents and the number
of approximate bits in the parent blocks. We demonstrate a significant improvement of accuracy in the prediction of
the noise power of DSP systems containing approximate adders. 
%We demonstrate as much as 6.5\,dB improvement in noise power prediction for 2D $8\times 8$ DCT implementation using an optimization framework.
\end{abstract}

\begin{IEEEkeywords}
Approximate computing, error model, static probability, optimization, noise power.
\end{IEEEkeywords}

\section{Introduction}
Approximate computing is  widely used in signal and image processing applications to obtain improvements in power and/or speed while
maintaining the required accuracy. Adders are the basic building blocks in these applications and a typical implementation has a large number of adders.
A variety of approximate adders have been proposed in the literature, with different levels of trade-offs between accuracy and
performance. These adders can be classified as low-latency \cite{GeAr2015} (and references therein) %\cite{acai2008,etaii2010,etaiii2011,ESA2011,GDA2013,acaiiDAC2012,GeAr2015} 
and low-power approximate adders (LPAA) %\cite{ErrormodelDAC2017}.
\cite{inexactadders2016,XORbased2013,AMA2013,LOA2010,zhu2010,CeliaDATE2018}.
Low power implementations of DSP systems using LPAAs optimize the power consumption by maximizing the number of approximate bits
in each adder for a given accuracy.
%our focus is power optimized implementations of signal processing algorithms using low power approximate adders (LPAA). This requires
%an optimization routine to find the maximum number of approximate bits possible in each adder for a given accuracy. This in turn
%requires accurate error models for the approximate adders.
%\textcolor{blue}{In the literature, multiple approaches have been proposed to obtain low power implementations of applications such as FIR filter \cite{FIR_newCAS2015,SABER_DAC2017,Framework_TCASI2019}, DCT \cite{JPEGDAC2016,Framework_TCASI2019,DCT_DATE2017} and FFT \cite{Framework_TCASI2019}. Low power implementations for these applications have been obtained by using a genetic programming-based method \cite{DCT_DATE2017}, or by replacing accurate building blocks (such as adders, multipliers) with approximate versions \cite{JPEGDAC2016,SABER_DAC2017,Framework_TCASI2019}.  The optimal configuration of approximate bits for different adders in the design can be obtained by performing exhaustive Monte-Carlo simulations~\cite{FIR_newCAS2015}, or by using an optimization framework that is based on error models for the approximate arithmetic blocks~\cite{JPEGDAC2016,SABER_DAC2017,DCT_DATE2017,Framework_TCASI2019}. Optimization algorithms based on mixed integer non-linear programming~\cite{JPEGDAC2016}, heuristics~\cite{SABER_DAC2017} and Lagrange multipliers~\cite{Framework_TCASI2019} have been proposed in the literature. }

In the literature, multiple approaches have been proposed to find optimal approximation levels for adders used in low power implementations. 
An approximate Finite Impulse Response (FIR) filter is designed by fixing the level of approximation of the adders using Monte-Carlo (MC) simulations in \cite{FIR_newCAS2015}. Approximate mirror adder-5 (AMA-5) \cite{AMA2013} modeled assuming uniformly distributed inputs are used in a 2D Discrete Cosine Transform (DCT) module constructed using 1D DCT blocks in \cite{JPEGDAC2016} and the optimization problem is solved using a mixed integer non-linear problem solver. Cartesian Genetic Programming (CGP) is used to design various approximate implementations of four point 1D DCT in \cite{DCT_DATE2017}. An expression for variance of error of AMA 1-5 adders \cite{AMA2013} and Lower-part-OR adder (LOA) \cite{LOA2010} is obtained in \cite{SABER_DAC2017} empirically by regression assuming uniform inputs, and heuristics are used to solve the approximation-level optimization problem. In \cite{Framework_TCASI2019}, AMA 1-5 adders and Transmission Gate based Approximate adders TGA I-II \cite{TGA2015} are considered. An expression for mean square error (MSE) is obtained assuming that the distribution of inputs and error are uniform. This is then used in a Lagrange multiplier based optimization approach. 

All of the above previous works on optimization use error metrics based on uniformly distributed inputs. Moreover, the same error model is used for all adders in the circuit. To verify the validity of these assumptions, we found the MSE in  an approximate $8\times 8$ DCT module \cite{BAS09}, that has
288 adders in a 6-level adder tree. We evaluated the noise power in dB ($10\times \log_{10} MSE$), assuming an error model for LOA based on uniformly distributed inputs \cite{LOA2010} and
compared it with values obtained using MC simulations, with $10^5$ uniformly distributed random inputs ($NP_1$ and $NP_{sim}$ respectively  in Table~\ref{tab:dct}). The numbers after levels ($L_1-L_6$) indicate the number of approximate bits. As seen from Table~\ref{tab:dct}, the analytical noise power differs from the simulated value by as much as 11\,dB, though the model worked well for an individual adder.

%we analytically compute the noise power ($NP_1$) at the output of an approximate $8\times 8$ DCT module \cite{BAS09} and compare it with that obtained using Monte-Carlo simulations ($NP_{sim}$) in Table~\ref{tab:dct}. For Monte-Carlo simulations, we considered $10^5$ uniformly distributed random inputs. The DCT module consists of 288 adders spread over 6 levels denoted by $L_1-L_6$ in the table. The accurate adders are replaced with LOAs. The inputs to the DCT module are assumed to have 15 bits of precision, i.e. Q1.15 in fixed point format. Analytical and simulated noise power are computed for various combinations of approximate bit assignments for adders in different levels as mentioned in the table. The absolute error in noise power prediction is given by $|e_1| = |NP_{sim}-NP_1|$.
%From Table~\ref{tab:dct}, we see that the analytical noise power differs from the simulated value by as much as 11\,dB.

While the assumption of uniformly distributed lower order bits may be justified for the primary inputs, neither the output of LPAAs nor the error  is uniformly distributed. A more accurate method of obtaining the probability mass function (PMF) of error is proposed in \cite{ErrorPMF_TCAD2019}.
However, %it is not practical to 
including this method within an optimization routine would require extensive computations. Moreover, in most applications, an accurate estimate of the mean error and MSE is sufficient and we do not need the PMF of the error. 
\setlength{\textfloatsep}{0.02cm}
\begin{table}[!tb]
\small
  \centering
  \caption{Noise power (in dB) of an $8\times 8$ DCT module that uses Lower part OR adders. $NP_1$: Noise power assuming uniform distribution, $NP_{sim}$: Noise power using simulations, $e=|NP_{sim}-NP_1|$.}
  \label{tab:dct}
\resizebox{\columnwidth}{!}{
  \begin{tabular}{|c|c|c|l|}
\hline
$NP_1$ &  $NP_{sim}$ & $e$  & No. of approximate bits\\
\hline
%-48.17  & -47.67 & 0.5  \\%&  1-6: 6\\
%\hline
-53.45 &  -51.20 & 2.25  & $L_1-L_4$: 5; $L_5-L_6$: 6\\
\hline
-51.76 & -44.67 & 7.09 & $L_1-L_2$: 5; $L_3-L_6$: 6\\
\hline
-51.32 & -43.00 & 8.32  & $L_1-L_2$: 5;  $L_3-L_5$: 6; $L_6$: 7\\
\hline
-50.55 & -41.21 & 9.34 & $L_1-L_2$: 5;  $L_3-L_4$: 6;  $L_5-L_6$: 7\\
\hline
-47.59 & -36.50 & 11.09 &  $L_1$: 5;  $L_2-L_3$: 6;  $L_4-L_5$: 7; $L_6$: 8\\
\hline
  \end{tabular}}
\end{table}\\

In this paper, we show that mean and MSE can be computed accurately without computing the PMF of the error
if the number of approximate bits and the static probabilities of the
inputs is taken into account. To this end,
we derive parameterised error models that can be used within any optimization framework in order to optimize the number of approximate bits.
We also show that in order to accurately compute the MSE, the optimization framework needs to take into account not just the functionality of
the adder, but also its position in the circuit, functionality of its parents and the number
of approximate bits in the parent blocks. We demonstrate a significant improvement of  accuracy in the prediction of MSE in applications
such as FIR filter, IIR filter and DCT using a variety of LPAAs.

Each approximate adder requires the static probabilities of its input bits, which eventually traces back to the primary inputs.
Even though the PMF of the input signal is usually very non-uniform, the PMF of the lower order bits is often close to uniform, so that
simple expressions for the error can be used.
Some justification for this assumption is included in \cite{CeliaDATE2018}, but it is heuristic and there is no condition that can be checked to test for
uniformity.
In this paper, we show that the discrete Fourier transform (DFT) of the input signal has to satisfy certain constraints if the PMF
is uniform. This can be used to easily find the maximum number of bits that can be considered to be uniformly distributed.

%check the validity of the assumption on primary input of an image processing application, where

%To summarize, our main contributions are as follows:
%\begin{enumerate}
%\item We show that the mean and MSE of the error can be predicted accurately if the stat
%\item We develop an optimization framework using parameterized error models for LPAAs to maximize the number
%of approximate bits in each adder of a signal processing application for a given accuracy constraint.
%\item We show that the Discrete Fourier Transform (DFT) of a signal satisfies a certain condition if the distribution of the lower order bits is uniform. 
%\item We obtain power-optimized implementations of an FIR filter and a 2D $8\times 8$ DCT module using the optimization framework and demonstrate significant improvement in accuracy prediction due to the use of the parameterized error models.
%\end{enumerate}

%We have organized this paper as follows: 
%In Section~\ref{Applications}, we present the methodology to derive the parameterized error model and to assign input probabilities to approximate adders in a multi-level circuit. In Section~\ref{qt}, we derive the conditions under which the lower order bits of the primary inputs will have uniform distribution.   
%The details of the optimizer that is developed to obtain the optimal number of bits that can be approximated for a given accuracy constraint are presented in Section~\ref{Optimizer}. 
%The results are shown in Section~\ref{Results} and finally Section~\ref{conclusion} concludes the paper.

%\input{background.tex}
\section{Parameterized error models for low power approximate adders}
\label{Applications}
As mentioned, the simple error models fail to give accurate estimates of the mean error and MSE since they typically assume that the inputs or
the error is uniformly distributed. The more complex models that compute the PMF of the error are more accurate, but unsuitable for use in
an optimizer. We propose parameterized error models with the input static probability as parameters, that can take into account the
input distribution without evaluating the full PMF. The models are more complex than the expressions in  \cite{Framework_TCASI2019}, but
they are analytical expressions, suitable for use in an optimizer.

Assume that an $(N,k)$-bit LPAA with inputs $a$ and $b$ has $k$
approximate bits and $N-k$ accurate bits, with $\hat{s}_i=f(a_i,b_i,\hat{c}_{i-1})$ and $\hat{c}_i=g(a_i,b_i,\hat{c}_{i-1})$  denoting the
$i^{th}$ bit of the approximate sum and carry. Let $P_{x_i}$ denote the static probability of a signal $x_i$. As with all models in the
literature, we continue to assume that $a$ and $b$ are independent. This assumption is an approximation when
the circuit has reconvergent fanouts. However, it is a reasonable approximation in many cases as correlations are diluted as the logic depth increases, as argued in \cite{ErrorPMF_TCAD2019}. 

The error in the output is due to the approximate lower part sum and the approximate carry to accurate adder. This can be written as 
\begin{align}
e = \sum\limits_{i=0}^{k-1} (a_i+b_i) 2^i - \sum\limits_{i=0}^{k-1} \hat{s}_i2^i - \hat{c}_{k-1} 2^k.\label{error}
\end{align}
The mean error of the approximate adder is given by 
\begin{align}
E\{e\} = \sum\limits_{i=0}^{k-1} (P_{a_i}+P_{b_i}) 2^i - \sum\limits_{i=0}^{k-1} P_{\hat{s}_i} 2^i - P_{\hat{c}_{k-1}} 2^k.\label{meanerror}
\end{align}
To derive $ P_{\hat{s}_i}$, we need $P_{\hat{c}_{i-1}}$. One way to do this is to start with the LSB, for which $P_{\hat{c}_{-1}} = 0$ and find the static
probability of all the other carry bits based on the truth table. Alternatively, we can assume that the static probability of the carry is
independent of the position i.e., $P_{\hat{c}_i} = P_{\hat{c}_{i-1}}$. This is a reasonable assumption as the input static probability is not expected
to vary with location. For example, for AMA-1 if we assume $P_{\hat{c}_i} = P_{\hat{c}_{i-1}}$, we get 0.67 as the static probability, whereas if we use
the first method, we get  [0.5,0.625,0.656,0.67,0.67] for the five LSBs as in \cite{AMA2013}. Using 0.67 as the static probability of the carry, $ P_{\hat{s}_i} = 0.33$ which
matches well with actual values for $k>3$.
%Once $P_{\hat{c}_i}$ is known,$ P_{\hat{s_i}}$ can also be computed using the truth table.
Note that we have not assumed anything about the PMF of the inputs. If it happens
to be uniform, we can use $P_{a_i}=0.5$. However, if the inputs come from another approximate adder, we just need to use the right value of the
static probability.

The MSE $(E\{e^2\})$ can be derived in a similar fashion. The expression for MSE computation requires evaluation of correlated terms such as $E\{\hat{s}_i\hat{s}_j\}$, $E\{a_i\hat{s}_j\}$, $E\{b_i\hat{s}_j\}$, $E\{a_i\hat{c}_{k-1}\}$, $E\{b_i\hat{c}_{k-1}\}$ and $E\{\hat{s}_i\hat{c}_{k-1}\}$ in terms of the static probabilities. Each of these can be derived using truth table of the approximate adder. The correlation between bits
$E\{\hat{s}_i\hat{s}_j\}$ can be neglected in most adders i.e., the individual bits of the sum are independent. However, it
is significant in a few adders such as AMA2 ($\hat{s}_i = \overline{\hat{c}_i}$), where there is a close relationship between the sum and carry.
An exception to this method for deriving error models is ETA-I \cite{zhu2010}, where the lower part sum cannot be written as a truth table.
Its error metrics are derived in our earlier work \cite{CeliaISCAS2018}.

\section{Multi-level systems}
In DSP systems, the inputs to the adder are either the primary inputs or they are output of another approximate adder, subtractor or (in our case, accurate) multiplier. In each case, the static probabilites of the input needs to be evaluated correctly. We consider each case in the following subsections.

\subsection{Static Probabilities: Primary inputs}
\label{qt}
Typical PMF of any primary input such as an image is not uniform. However, since error expression of a LPAA involves the static probability of the lower $k$ bits, we would only need to check if the PMF of these $k$ bits is uniform.
% In a previous work \cite{CeliaDATE2018}, we gave some justification for this assumption, but it was not an exact condition that could be checked.
A brute-force technique to check would be to compute the PMF of the lower order bits for each value of $k$ and check for uniformity of the PMF. In this  section, we show  that it is possible to check if the distribution is uniform by computing the DFT of the input. A  single DFT  is  sufficient to  find  the  values of $k$ for which this assumption is reasonable.

%Typical PMF of any primary input such as an image is not uniform. %As an example, the PMF of Cameraman image is shown in Fig.~\ref{imagePMF}. However, we are concerned about the PMF of the lower bits of the input signal or image, since error expression of a LPAA involves the probability of the lower $k$ bits that are approximated. In all the previous works, the distribution of lower order bits of all primary inputs are assumed to be uniform. 
%In \cite{Qnoise1977}, the authors derive the condition under which the quantization error can be considered to be uniform, when a continuous time signal is input to a quantizer. 
%We now derive conditions for the $k$ LSBs of an $N$ bit signal to be uniform. 
\noindent %$P_A$: PMF of the $N$-bit input, $A$ \\

Let $\mathcal{F}_{A}$ be the $2^N$-point DFT of the PMF of $N$-bit signal $A$ and $\mathcal{F}_{A_L}$ be the $2^k$-point DFT of the PMF of $A_L$ ($k$ LSBs of $A$). We have, 
\begin{align}
\mathcal{F}_{A_L}[m] =& \sum\limits_{n=0}^{2^k-1} P(A_L=n)e^{-jmn2\pi/2^k} \label{eqn:phiQ}\\
=& \sum\limits_{n=0}^{2^k-1} \sum\limits_{l=0}^{2^{N-k}-1} P(A=l2^k+n) e^{-jmn2\pi/2^k} \nonumber \\
%=& \sum\limits_{l=0}^{2^{N-k}-1} \sum\limits_{n'=l2^k}^{l2^k+2^k-1} P(A=n')e^{-jm2\pi/2^k  (n'-l2^k)} \nonumber \\
=& \sum\limits_{l=0}^{2^{N-k}-1} \sum\limits_{n'=l2^k}^{l2^k+2^k-1} P(A=n')e^{-jmn'2\pi/2^k} e^{jml2^k2\pi/2^k } \nonumber\\
=&\sum\limits_{n'=0}^{2^N-1} P(A=n')e^{-jmn'2\pi/2^k} \nonumber\\
=& \mathcal{F}_A[m\cdot 2^{N-k}], 0\le m<2^k.\label{phiQeqA}
\end{align}
%since we know that 
%\begin{align}
%\mathcal{F}\{A\} =& \sum\limits_{n=0}^{2^N-1} P(A=n)e^{jm2\pi/2^N n}. \label{eqn:phiA}
%\end{align}
If $A_L$ is uniform, %\begin{align}
$P(A_L=n)= \dfrac{1}{2^k} , 0\le n<2^k$. 
Hence from \eqref{eqn:phiQ}, if $A_L$ is uniform, we have
\begin{align}
\mathcal{F}_{A_L}[m] =& \sum\limits_{n=0}^{2^k-1}\dfrac{1}{2^k} e^{-jmn2\pi/2^k} = 
\begin{cases}
   1,        & \text{if }m=0 \\
   0,        & \text{if }0<m<2^k.
  \end{cases}\label{CFofQcondition}
\end{align}
Since DFT is unique, the converse is also true. %if the transform satisfies this condition, the lower $k$ bits are uniformly distributed. 
Therefore using \eqref{phiQeqA}, we have the following condition to be satisfied for $A_L$ to be uniformly distributed.
\begin{align}
\mathcal{F}_A[m\cdot 2^{N-k}] &= 
\begin{cases}
   1,        & \text{if }m=0 \\
   0,        & \text{if }0<m<2^k.
  \end{cases}\label{CFcondition}
\end{align}
In \cite{Qnoise1977}, they have similar condition for continuous signals that are quantized, although the derivation is a little more involved.  

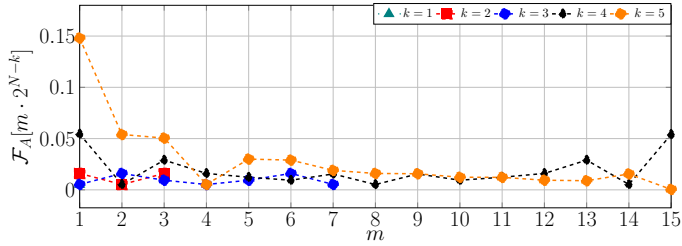
\begin{figure}[!b]
\begin{tikzpicture}[scale=0.475]
\begin{axis}[
legend columns = 5,
	scaled y ticks = false,
	y tick label style={/pgf/number format/fixed},
	height=0.4\textwidth,
	width=1.0\textwidth,
label style={font=\LARGE},
     tick label style={font=\LARGE} ,
y label style={at={(-0.025,0.5)}},
     xlabel= $m$,
     ylabel= $\mathcal{F}_A{[}m\cdot 2^{N-k}{]}$,
	ymax = 0.18,
	xmin = 1,
	xmax = 15,
	grid=major,
legend style={at={(1,1.01)},anchor=north east}
]
%k=1
\addplot [ycomb, teal, very thick, mark=triangle*, mark size =4]%, mark options={blue!60!black}]
coordinates {
( 1 ,  0.0052 )
};
%k=2
\addplot [dashed, red, very thick, mark=square*, mark size =4]%, mark options={red!60!black}]
coordinates {
( 1 ,  0.016 )
( 2 ,  0.0052 )
( 3 , 0.016 )
};
%k=3
\addplot [dashed, blue, very thick, mark=*, mark size =4]%, mark options={teal!60!black}]
coordinates {
( 1 ,  0.0053 )
( 2 ,  0.016 )
( 3 ,  0.0094 )
( 4 ,  0.0052 )
( 5 ,  0.0094 )
( 6 ,  0.016 )
( 7 ,  0.0053 )
};
%k=4
\addplot [dashed, black, very thick, mark=diamond*, mark size=4]%, mark options={black!60!black}]
coordinates {
( 1 ,  0.0544 )
( 2 ,  0.0053 )
( 3 ,  0.029 )
( 4 ,  0.016 )
( 5 ,  0.0123 )
( 6 ,  0.0094 )
( 7 ,  0.0155 )
( 8 , 0.0052 )
( 9 ,  0.0155 )
( 10 , 0.0094 )
( 11 ,  0.0123 )
( 12 , 0.016 )
( 13 ,  0.029 )
( 14 , 0.0053 )
( 15 , 0.054 )
};
%k=5
\addplot [dashed, orange, very thick, mark=*, mark size =4]%, mark options={orange!60!black}]
coordinates {
( 1 ,  0.148 )
( 2 ,  0.054 )
( 3 ,  0.0505 )
( 4 ,  0.0053 )
( 5 ,  0.03 )
( 6 ,  0.029 )
( 7 ,  0.019 )
( 8 , 0.016 )
( 9 ,  0.0157 )
( 10 , 0.0123 )
( 11 ,  0.0122 )
( 12 , 0.0094 )
( 13 ,  0.0089 )
( 14 , 0.0155 )
( 15 , 0.00043 )
};
\legend{$k=1$,$k=2$,$k=3$,$k=4$,$k=5$}
\end{axis}
\end{tikzpicture}
\caption{Illustration of condition for $k$ lower-order bits of Cameraman image to be uniform.}
\label{charFunction}
\end{figure}
\input{fig-histLSB.tex}

To illustrate this condition \eqref{CFcondition}, let us consider the Cameraman image with $N=8$. For the image pixel distribution, $\mathcal{F}_A[m\cdot 2^{N-k}]$ for different values of $k$, $m$ varying from $0$ to $2^k-1$, is plotted in Fig.~\ref{charFunction}. It is seen that for lower values of $k$, the value of the transform is very close to zero for $0<m<2^k$. As $k$ increases, the value of transform also increases and for $k=5$, the values are high.
This is confirmed from the actual PMF of the lower order bits of the image shown in Fig.~\ref{fig:histLSB}.
From the figure, it is seen that distribution can be considered uniform even if half the bits are approximated. This turns out to be true for all
the standard images we have looked at (for example, for the Lena image, $k=6$, for Rice, $k=5$). Hence, we assume that the PMF of the lower order bits of primary inputs to the approximate adder is uniform i.e., the static probability is 0.5.

%%*************************************************************************************
\subsection{Static Probabilities: Adders in the higher levels}
\label{piAdder}
%The two operations used in a linear time-invariant signal processing system are addition of two inputs and multiplication with a constant.

\begin{figure}[!tb]
  \center
 \begin{subfigure}{0.48\columnwidth}
\begin{tikzpicture}[thick,scale=0.75, every node/.style={scale=0.75}]
\draw [ thick](0,0) circle (0.25cm) node {$\boldsymbol{+}$};
\draw [ thick](2,0) circle (0.25cm) node {$\boldsymbol{+}$};
\draw [ thick](1,-1.5) circle (0.25cm) node {$\boldsymbol{+}$};

\draw[->, thick] (-0.5,0.75)--(-0.15,0.2) ;
\draw[->, thick] (0.5,0.75)--(0.15,0.2) ;
\draw[->, thick] (1.5,0.75)--(1.85,0.2) ;
\draw[->, thick] (2.5,0.75)--(2.15,0.2) ;

\draw[->, thick] (0,-0.25)--(0.8,-1.35) ;
\draw[->, thick] (2,-0.25)--(1.2,-1.35) ;

\draw[->, thick] (1,-1.75)--(1,-2.5) ;

\draw[.] (-0.35,0.5) node[left]{\textbf{$N$}};
\draw[.] (0.3,0.5) node[right]{\textbf{$N$}};
\draw[.] (1.65,0.5) node[left]{\textbf{$N$}};
\draw[.] (2.3,0.5) node[right]{\textbf{$N$}};
\draw[.] (0,-0.75) node[left]{\textbf{$m_1,$}};
\draw[.] (0.35,-0.75) node[left] {\textcolor{red}{\textbf{$k_1$}}};
\draw[.] (1.75,-0.75) node[right]{\textbf{$m_2,$}};%{\textcolor{blue}{\textbf{$k_2$}}};
\draw[.] (2.35,-0.75) node[right] {\textcolor{red}{\textbf{$k_2$}}};
\draw[.] (1,-2.25) node[right]{\textbf{$m_3,$}};%{\textcolor{blue}{\textbf{$k_2$}}};
\draw[.] (1.6,-2.25) node[right] {\textcolor{red}{\textbf{$k_3$}}};

\draw[.] (-0.25,0) node[left]{\textbf{Adder1}};
\draw[.] (2.25,0) node[right]{\textbf{Adder2}};
\draw[.] (-0.75,-1.5) node[right]{\textbf{Adder3}};

\end{tikzpicture}
\subcaption{}
  \label{fig:addertree}
\end{subfigure}
\begin{subfigure}{0.48\columnwidth}
 \begin{tikzpicture}[thick,scale=0.8, every node/.style={scale=0.8}]
\draw [ thick](0,0) circle (0.25cm) node {$\boldsymbol{+}$};
\draw [ thick](2,0) circle (0.25cm) node {$\boldsymbol{+}$};
\draw [ thick](0,-1.5) circle (0.25cm) node {$\boldsymbol{+}$};
\draw [ thick](1.1,-0.9) circle (0.25cm) node {$\boldsymbol{*}$};

\draw[->, thick] (-0.5,0.75)--(-0.15,0.2) ;
\draw[->, thick] (0.5,0.75)--(0.15,0.2) ;
\draw[->, thick] (1.5,0.75)--(1.85,0.2) ;
\draw[->, thick] (2.5,0.75)--(2.15,0.2) ;

\draw[->, thick] (0,-0.25)--(0,-1.25) ;
\draw[->, thick] (2,-0.25)--(1.3,-0.8) ;
\draw[->, thick] (0.9,-1.)--(0.25,-1.5) ;
\draw[->, thick] (0.65,-0.4)--(0.9,-0.7) ;

\draw[->, thick] (0,-1.75)--(0,-2.5) ;

\draw[.] (-0.35,0.5) node[left]{\textbf{$N$}};
\draw[.] (0.3,0.5) node[right]{\textbf{$N$}};
\draw[.] (1.65,0.5) node[left]{\textbf{$N$}};
\draw[.] (2.3,0.5) node[right]{\textbf{$N$}};
\draw[.] (-0.25,-0.55) node[left]{\textbf{$m_1,$}};
\draw[.] (0.1,-0.55) node[left] {\textcolor{red}{\textbf{$k_1$}}};
\draw[.] (-0.05,-0.95) node[left] {\textcolor{blue}{\textbf{$(P_{a_i})$}}};
\draw[.] (1.65,-0.65) node[right]{\textbf{$m_2,$}};%{\textcolor{blue}{\textbf{$k_2$}}};
\draw[.] (2.5,-0.95) node[left]{\textcolor{blue}{\textbf{$(P_i)$}}};
\draw[.] (1.5,-1.35) node[left]{\textcolor{blue}{\textbf{$(P_{b_i})$}}};
\draw[.] (2.25,-0.65) node[right] {\textcolor{red}{\textbf{$k_2$}}};
\draw[.] (0.2,-2.25) node[right]{\textbf{$m_3,$}};%{\textcolor{blue}{\textbf{$k_2$}}};
\draw[.] (0.8,-2.25) node[right] {\textcolor{red}{\textbf{$k_3$}}};

\draw[.] (0.65,-0.45) node[above]{c};
\draw[.] (-0.25,0) node[left]{\textbf{Adder1}};
\draw[.] (2.25,0) node[right]{\textbf{Adder2}};
\draw[.] (-0.25,-1.5) node[left]{\textbf{Adder3}};

\end{tikzpicture}
\subcaption{}
  \label{fig:addermulttree}
\end{subfigure}

  \caption{Adder tree in a circuit with $N$ bits at primary inputs.}
  \label{fig:tree}
\end{figure}
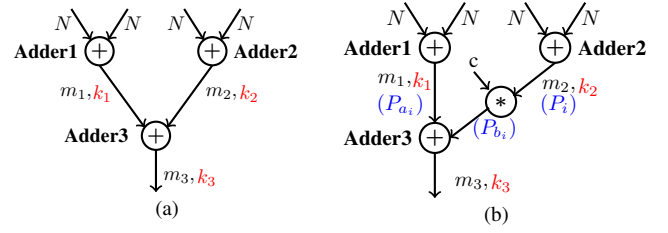

%First, let us consider the case that the adder input comes from another adder's output. For example, consider Adder3 in Fig.~\ref{fig:addermulttree}, with one of the inputs from Adder1. The corresponding input bit probability $P_{a_i}$ depends upon the Boolean expression used for the computation of its lower part sum. 
%For example, in case of Truncation adder, the LSBs are fixed to constant 0's. i.e. $P_{a_i}=0$ for $k_1$ LSBs. 
If the inputs to the adder are the output of another adder as in Fig.~\ref{fig:addertree}, the mean and MSE are derived using
$P_{\hat{s}_i}$ and $P_{\hat{c}_{k-1}}$ as discussed previously.

%%*************************************************************************************
The other possibility is that input is the output of a  multiplier. In this work, we are only optimizing adders and all multipliers are accurate,  with the output truncated to the standard precision used in the circuit. Also, we only consider linear systems, so that one of the inputs to the
multiplier is a constant coefficient. 
In Fig.~\ref{fig:addermulttree}, consider Adder3 which has an input from the output of the multiplier. Depending on the value of the constant coefficient $c$, the probability of the LSBs at the output of the multiplier ($P_{b_i}$) will vary. Let $P_{i}$ denote the probability that the $i^{th}$ bit of the $k_2$ LSBs of the multiplicand (output of Adder2) is 1. 
Consider the following cases. 
\begin{enumerate}
\item When $c=2^l$ and $l\ge 0$, the product is the logical left shift of the multiplicand. So $P_{b_i}=0$ for the first $l$ LSBs and $P_{b_i}=P_{i-l}$ for the next $k_2-l$ LSBs.
\item When $c=-2^l$ and $l\ge 0$, $P_{b_i}=0$ for the first $l$ LSBs and $P_{b_i}=1-P_{i-l}$ for the next $k_2-l$ LSBs is a good approximation, accounting for flipping involved in the two's complement representation for negative numbers.
\item When $c=2^l$ and $l<0$, the product is the right shift of the multiplicand. So  $P_{b_i}=P_{i+|l|}$ for $k_2-l$ LSBs.
\item When $c=-2^l$ and $l<0$, $P_{b_i}=1-P_{i+|l|}$ for $k_2-l$ LSBs.
\item If $c$ is an arbitrary constant that is not a power of 2, MC simulations indicate that the average static probability of the output bits is 
  $0.5\pm 0.03$. This is also intuitively correct, since the product is the sum of several partial products and the LSBs are
  truncated to the precision maintained in the system. Hence, the overall PMF is likely to be symmetric (moving towards Gaussian), which means that the static probability is
  %once again 
0.5. So
  %Therefore, in this case, 
we assume that $P_{b_i}=0.5$ for $k_2$ LSBs. 
%\item For $c$ chosen uniformly at random, Monte Carlo simulations indicate that the average static probability of the output bits is $0.5\pm 0.03$ for each of the LSBs. Therefore, when $c$ is not a power of 2, we assume that $P_{b_i}=0.5$ for $k_2$ LSBs. 
\end{enumerate}

%%*************************************************************************************
\subsection{Truncation and Median adder (MA) in higher levels}
\label{carryMA}
Both Truncation and MA \cite{CeliaDATE2018} have their lower part sum bits fixed to constant all 0's and 1's respectively. In these adders, since the lower part sum is known, the lower part sum of the adders in higher levels can be fixed more accurately so that the accuracy of the approximate circuit is improved.
%For example, consider an adder tree as shown in Fig.~\ref{fig:addertree} with Truncation adders. Since the lower part sum of Adder1 and Adder2 are zero, even the accurate lower part sum of Adder3 is zero. In this case, the accuracy can not be improved further. 
In case of Truncation adder, the approximate sum is obviously zero. In Fig.~\ref{fig:addertree} with MA,
Adder1 and Adder2 will have their lower part sum as $2^{k_1}-1$ and $2^{k_2}-1$, respectively.
For Adder3, %instead of setting the lower part sum as $2^{k_3}-1$, %($\hat{c}_{k-1}=0$, $\hat{s}_i=1$),  
we lower the MSE of the circuit by setting the sum to $2^{k_3+1}-1$ instead of $2^{k_3}-1$ %($\hat{c}_{k-1}=1$, $\hat{s}_i=1$) 
for the following cases: 
\begin{enumerate}
\item When $k_3 \le k_1,k_2$, the lower part sum is known exactly and is equal to $2^{k_3+1}-2$. %, which is closer to $2^{k_3+1}\!-\!1$ than $2^{k_3}\!-\!1$. 
\item When $k_1\ge k_3>k_2$, the mean of the sum is $(3\times 2^{k_3}+2^{k_2}-4)/2$, which is closer to $2^{k_3+1}-1$ than $2^{k_3}-1$. 
\end{enumerate}
Using this setting, we obtain up to 6\,dB improvement in MSE for the adder tree in Fig.~\ref{fig:addertree}. Beyond the first level, the input static probability for Median adders is set to 1.
%%*************************************************************************************

\section{Optimization framework}
\label{Optimizer}
Classical wordlength optimization uses a simple model for the quantization error and the same model is used for all nodes in the circuit.
In the literature, a similar framework with a single expression for error for all the adders has been used to find the optimal number of approximate bits. However, the discussion in the previous
section shows that this is inadequate to get accurate numbers. Hence, we made several modifications to the framework, which are detailed below.

The input to the optimizer is the circuit implemented using adders, multipliers and registers and the corresponding signal flow graph.
 %The overall MSE is computed based on the mean error, MSE and the transfer function from each adder to the output. 
%Given the block diagram of a circuit, we construct a directed acyclic graph (DAG) with the functional elements such as adders, multipliers and registers as nodes and connections between them as edges. 
The primary inputs to the system are normalized to $1.N$ fixed point numbers.
For each functional unit in the system, we use the required number of integer bits while maintaining the number of fractional bits as $N$.
The goal of the optimizer is to maximize the number of approximate fractional bits of the adders in the circuit for a given MSE at the output.  
We use the three step procedure discussed in \cite{WLO} and adapt it for approximate computing. It uses
Minimum Width algorithm, Mildest Greedy Ascent algorithm and Tabu search algorithm. The main steps in our optimization framework are as follows:
%was used to minimize the word length of each signal in the circuit for a given quantization noise constraint. We use a similar three-step procedure to maximize the number of approximate bits in each adder, for a given MSE constraint.  Several changes were introduced in the optimization for this purpose and the main steps are described below. %This requires the incorporate %For this purpose, several modifications were made in the algorithms in order to incorporate the parameterized   %simultaneous satisfaction of constraints at multiple outputs.
%The main steps in our optimization framework are as follows:
\begin{itemize}
%\item In \cite{WLO}, the error introduced at each node due to quantization depended only on the number of bits quantized at that node. In our case, we need to keep track of the number of approximate bits in each adder, its parent nodes and their functionality and the type of approximate adder. 
\item We have a pre-processing step in which adder gets the static probability of the inputs based on its parent nodes. If the parent is a register, it
  gets the static probability of the inputs to the register. Also, the transfer function from each adder node to the output is computed. 
  %We incorporate the parameterized error models of various approximate adders to compute the MSE introduced at each adder node due to approximation. The input static probabilities of each adder node are computed as part of a pre-processing step using the DAG as described in Section~\ref{piAdder}. %The overall MSE at the output can be computed using the transfer functions from the adder nodes. 
\item Next, we run a minimum width algorithm (MWA) that gives the maximum number of approximate fractional bits at the output of each adder when all the other adders are accurate and the required MSE constraint is satisfied.
\item Starting with the number of approximate bits from the MWA, a greedy descent procedure is used to decrease the number of approximate bits in the adder that causes the maximum improvement in MSE. An important difference from the quantization noise optimization in \cite{WLO} is that the approximation noise can worsen even if the number of approximate bits is decreased. % in certain cases.  
%. Although counter-intuitive, this happens as the mean error shifts significantly for some of the adders.
For circuits with multiple outputs, we find the fan-in cone of each output in sequence. The adders in the fan-in cone are optimized while keeping the adders in the fan-in cone that increase the MSE of previously targeted outputs untouched.
\item Finally we run a tabu search algorithm targeting signals with minimum number of approximate bits (instead of the most sensitive signal) and keep increasing the number of approximate bits as long as MSE constraint is met. We found that this heuristic provides better optimization of approximate bits. 
\end{itemize}
In each of these algorithms, the overall MSE at the output is computed using the transfer functions from the adder nodes and the parameterized error model for the adder. 

\section{Experimental Results}
\label{Results}

In this section, we first validate our error model. Each assumption is tested against simulation and verified for correctness. We then obtain the
optimum number of approximate bits for a given MSE using our optimization framework. This is done for FIR and IIR filters and an $8\times 8$ DCT module.
These results are used to show that the optimizer requires the parameterized error models with the correct values of the static probability
for accurate prediction of the MSE.% (also referred to as the noise power). 
%%%%%%%%%%%%%%%%%%%%%%%%%%%%%%%%%%%%%%%%%%%%%%%%%%%%%%%%%%%%%%%%%%%%%%%%%%%%%%%%%%%%%%%%%%%%%
%\subsection{ Comparison of the analytical and simulated value of MSE }

In order to validate our error model, we used the simple adder tree depicted in Fig.~\ref{fig:addertree}. 
%Table~\ref{loa-np-tree-adderi} shows
%a comparison of the simulated and analytical values of the noise power for Adder3 in the LOA adder tree. The row $NP_{sim}$ shows the MC simulation results for various number of approximate bits $k$. The rows $NP_{p_i=0.5}$ and $NP_{p_i=0.75}$ give the values computed analytically assuming that the input static probability of Adder3 as 0.5 and 0.75 respectively. It is seen from the table that accurate prediction of the MSE requires the correct value of the static probability.
%Moreover, the mean error is also significantly different with  $\mu_{p_i=0.5}=-0.25$ as opposed to $-0.5625$ obtained using simulations and
%analytical expressions with the correct value of the static probability.
%\input{fig/tab-loa-np-tree-adderi.tex}
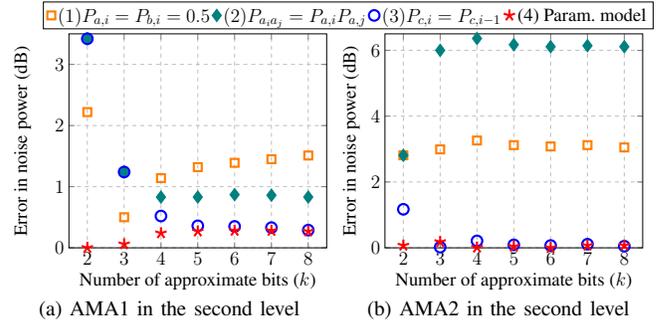
\begin{figure}[!tb]
\begin{subfigure}[b]{0.48\columnwidth}
\begin{tikzpicture}[scale=0.5]
   \begin{axis}[
legend cell align={left},
legend columns = 4,
every axis plot/.append style={ultra thick},
     xmin=1.5, xmax=8.5,
     ymin=0, ymax=3.5,
     xlabel= Number of approximate bits ($k$),
     ylabel= Error in noise power (dB),
label style={font=\Large},
y label style={at={(0,0.5)}},
                    tick label style={font=\Large} ,
    ymajorgrids=true,
    grid style=dashed,
	grid=major,
	legend style={font=\Large,at={(-0.1,1.15)},anchor=north west, legend columns = 3}]

%MSE3_paipbi05
\addplot [color=orange, mark=square, mark size=1.5,only marks,  mark options={scale=2,solid}] coordinates {
(2, 2.22)
(3, 0.5)
(4, 1.14)
(5, 1.32)
(6, 1.39)
(7, 1.45)
(8, 1.51)
};

%MSE3_sisj_independent
\addplot [color=teal, mark=diamond*, mark size=2,only marks,  mark options={scale=2,solid}] coordinates {
(2, 3.42)
(3, 1.24)
(4, 0.83)
(5, 0.83)
(6, 0.87)
(7, 0.86)
(8, 0.83)
};
%MSE3_anal and param model with pci=pci-1
\addplot [only marks, color=blue, mark=o, mark size=2, mark options={scale=2,solid}] coordinates {
(2, 3.42)
(3, 1.24)
(4, 0.52)
(5, 0.36)
(6, 0.35)
(7, 0.33)
(8, 0.29)
};

%MSE3_anal and param model wo pci=pci-1
\addplot [only marks, color=red, mark=star, mark size=2.25, mark options={scale=2,solid}] coordinates {
(2, 0)
(3, 0.06)
(4, 0.24)
(5, 0.27)
(6, 0.28)
(7, 0.28)
(8, 0.27)
};
\legend{%uniform error\cite{Framework_TCASI2019}, 
$(1) P_{a,i}=P_{b,i}=0.5$, $(2) P_{a_ia_j}=P_{a,i}P_{a,j}$, $(3) P_{c,i}=P_{c,i-1}$, (4) Param. model}
\end{axis}
\end{tikzpicture}
\subcaption{AMA1 in the second level}
  \label{fig:ama1-mse3}
\end{subfigure}
%\subcaption{Variation in Mean Error Distance with the number of approximate bits in various 16-bit approximate adders}
\begin{subfigure}[b]{0.48\columnwidth}
\begin{tikzpicture}[scale=0.5]
   \begin{axis}[
legend cell align={left},
legend columns = 1,
every axis plot/.append style={ultra thick},
     xmin=1.5, xmax=8.5,
     ymin=0, ymax=6.5,
     xlabel= Number of approximate bits ($k$),
     ylabel= Error in noise power (dB),
label style={font=\Large},
y label style={at={(0.05,0.5)}},
                    tick label style={font=\Large} ,
    ymajorgrids=true,
    grid style=dashed,
	grid=major,
	legend style={font=\Large,at={(0,1)},anchor=north west}] 

%MSE3_paipbi05
\addplot [color=orange, mark=square, mark size=1.5,only marks, mark options={scale=2,solid}] coordinates {
(2, 2.81)
(3, 2.99)
(4, 3.26)
(5, 3.12)
(6, 3.08)
(7, 3.12)
(8, 3.05)
};

%MSE3_sisj_independent
\addplot [color=teal, mark=diamond*, mark size=2,only marks, mark options={scale=2,solid}] coordinates {
(2, 2.81)
(3, 6)
(4, 6.36)
(5, 6.17)
(6, 6.11)
(7, 6.14)
(8, 6.11)
};

%MSE3_anal assuming pci=pci-1 and param model
\addplot [only marks, color=blue, mark=o, mark size=2, mark options={scale=2,solid}] coordinates {
(2, 1.17)
(3, 0.02)
(4, 0.21)
(5, 0.09)
(6, 0.07)
(7, 0.11)
(8, 0.05)
};
%MSE3_anal and param model wo pci=pci-1
\addplot [only marks, color=red, mark=star, mark size=2.25, mark options={scale=2,solid}] coordinates {
(2, 0.07)
(3, 0.18)
(4, 0.01)
(5, 0.04)
(6, 0)
(7, 0.07)
(8, 0.07)
};

%\legend{%uniform error\cite{Framework_TCASI2019}, 
%$(1) P_{a,i}=P_{b,i}=0.5$, $(2) P_{s_is_j}=P_{s,i}P_{s,j}$, (3) Param. model}
\end{axis}
\end{tikzpicture}
\subcaption{AMA2 in the second level}
  \label{fig:ama2-mse3}
\end{subfigure}
  \caption{Error in noise power computation of Adder3 in Fig.\ref{fig:addertree} when the tree is implemented using  (a) AMA-1 and (b) AMA-2 adders.}
\label{fig:ama12addertree_msei}  
\end{figure}
Some of the LPAAs like AMA-1 and AMA-2 adders involve carry propagation in the approximate lower part sum. Therefore, in addition to the correct value of the static probability, evaluation
of the MSE also requires correlations between bits to be taken into account.  Fig.~\ref{fig:ama12addertree_msei} shows a comparison of the error in noise power computation
 of Adder3 in Fig.~\ref{fig:addertree} %(a) assuming uniform error as in \cite{Framework_TCASI2019} 
(1) assuming $P_{a,i}=P_{b,i}=0.5$ (2) assuming that the individual bits of each input are independent (i.e. $P_{a_ia_j}=P_{a,i}P_{a_j}$)  (3) assuming that $P_{c,i}=P_{c,i-1}$ (4) using the parameterized error model including all the correlations.
It can be seen that, $P_{c,i}=P_{c,i-1}$ is a good approximation for $k>3$. The correlations between the bits in each input 
can be ignored in AMA-1, AMA-2 requires all correlations to be taken into account. From the discussion in Section~\ref{Applications}, both these results are as expected.

%%%%%%%%%%%%%%%%%%%%%%%%%%%%%%%%%%%%%%%%%%%%%%%%%%%%%%%%%%%%%%%%%%%%%%%%%%%%%%%%%%%%%%%%%%%%%

%\subsection{Validation of Optimization framework}

We validated our optimization framework using an 18-tap FIR filter (direct form~I realization), IIR filter (direct form~II realization of a 4\ts{th} order low pass Butterworth filter) % obtained by cascading two second order filters)
 and DCT \cite{BAS09} consisting of 17, 8 and 288 adders respectively. These are typically the benchmarks that have
been used in the literature. For each of these systems, we obtained the optimum number of approximate bits for each adder in the system,  
given an overall MSE specification. This was done for the Truncation, MA, AMA-5, LOA and ETA-I adders. These adders have minimal hardware for evaluation of the approximate sum and are energy-efficient. Using the optimal configuration of approximate bits for each adder obtained from the optimizer, the circuits are implemented and simulated with $10^5$ uniform random inputs to obtain the actual MSE of the circuits. The actual MSE is compared with the analytical value obtained using the parameterized error model.

Figs.~\ref{fig:sim_anal_fir} and \ref{fig:sim_anal_iir} show the comparison of simulated value of noise power % in dB, which is $10\times \log_{10} MSE$, 
and analytically computed noise power in dB for the FIR filter and IIR filter respectively using various approximate adders for two cases - (a) assuming that the input static probabilities are 0.5 for all the adders (b) %using the proposed error model by
computing the input static probabilities for adders %in the higher levels of the circuit 
as described in Section~\ref{piAdder}. It can be seen that the analytical values obtained by using the parameterized error model match very well with the actual values, while those obtained using uniform probabilities deviate significantly (upto 6\,dB and 10\,dB error in case of FIR and IIR filters respectively).
\begin{figure}[!tb]
\begin{subfigure}{0.48\columnwidth}
\begin{tikzpicture}[scale=0.525]
\begin{axis}[
legend columns = 3,
every axis plot/.append style={ultra thick},
y label style={at={(0.05,0.5)}},
     xmin=-65, xmax=-20,
     ymin=-65, ymax=-20,
     xlabel= Actual noise power (dB),
     ylabel= Analytical noise power (dB),
label style={font=\large},
                    tick label style={font=\small} ,
    ymajorgrids=true,
    grid style=dashed,
	grid=major,
	legend style={font=\small,at={(0,1.0)},anchor=north west}] %($\mu$ W)]
%Trunc_pi05
\addplot [dashed,mark options={scale=1,solid},color=cyan, mark=square, mark size=1.5] coordinates {
( -30.6 ,  -30.4 )
( -35.4 ,  -35.06 )
( -40.75 ,  -40.09 )
( -46.25 ,  -45.03 )
( -52.54 ,  -50.17 )
( -59.8, -55.07)
};

%AMA5
\addplot [dashed,mark options={scale=1,solid}, color=orange,mark=*, mark size=1, mark options={scale=2,solid}] coordinates {
( -30.23 ,  -30.19 )
( -35.09 ,  -35.01 )
( -40.0 ,  -40.08 )
( -45.35 , -45.35 )
( -50.33 , -50.22 )
(-55.45 , -55.14)
};

%LOA
\addplot [dashed,mark options={scale=1,solid}, color=teal, mark=star, mark size=1.75, mark options={scale=2,solid}] coordinates {
( -28.4,  -30.03 )
( -33.07 ,  -35.15 )
( -35.7 ,  -40.02 )
( -40.46 ,  -45.09 )
( -45.72 ,  -50.23 )
( -52.7 ,  -55.16 )
( -58.7 ,  -60.03 )
};

%ETA
\addplot [dashed,mark options={scale=1,solid}, color=black, mark=o, mark size=1, mark options={scale=2,solid}] coordinates {
( -23.48 ,  -30.02 )
( -28.72 ,  -35.0 )
( -34.45 ,  -40.12 )
( -39.78 ,  -45.06 )
( -46.0 ,  -50.04 )
( -53.7 ,  -55.02 )
%( -68.25 ,  -60.06 )
};

%MA1
\addplot [dashed,mark options={scale=1,solid},color=red, mark=diamond*, mark size=2.5] coordinates {
( -28 ,  -30.29 )
( -31.4 , -35.29 )
( -36.2 , -40.12 )
( -43.1 , -45.16 )
( -52.37 , -50.0 )
( -58.26 , -55 )
( -62.23 , -60.08 )
};

\legend{Trunc,AMA5,LOA,ETA-I,MA}
\end{axis}
\end{tikzpicture}
\subcaption{Uniform distribution}
  \label{fig:WL_fir}
\end{subfigure}
\begin{subfigure}{0.48\columnwidth}
  \centering
\begin{tikzpicture}[scale=0.525]
\begin{axis}[
legend columns = 3,
every axis plot/.append style={ultra thick},
y label style={at={(0.05,0.5)}},
     xmin=-65, xmax=-20,
     ymin=-65, ymax=-20,
     xlabel= Actual noise power (dB),
     ylabel= Analytical noise power (dB),
label style={font=\large},
                    tick label style={font=\small} ,
    ymajorgrids=true,
    grid style=dashed,
	grid=major,
	legend style={font=\small,at={(0,1.0)},anchor=north west}] %($\mu$ W)]

%Trunc_pi0
\addplot [dashed,mark options={scale=1,solid},color=cyan, mark=square, mark size=1.5] coordinates {
( -30.6 ,  -30.4 )
( -35.4 ,  -35.06 )
( -40.75 ,  -40.09 )
( -46.25 ,  -45.03 )
( -52.54 ,  -50.17 )
( -59.8, -55.07)
};

%AMA5
\addplot [dashed,mark options={scale=1,solid}, color=orange,mark=*, mark size=1, mark options={scale=2,solid}] coordinates {
( -30.23 ,  -30.19 )
( -35.09 ,  -35.01 )
( -40.0 ,  -40.08 )
( -45.35 , -45.35 )
( -50.33 , -50.22 )
( -55.45 , -55.14)
};
%LOA
\addplot [dashed,mark options={scale=1,solid}, color=teal, mark=star, mark size=1.75, mark options={scale=2,solid}] coordinates {
( -29.26 ,  -30.02 )
( -34.39 ,  -35.26 )
( -39.7 ,  -40.1 )
( -45.17 ,  -45.17 )
( -50.43 ,  -50.0 )
( -55.92 ,  -55.07 )
( -61.83 ,  -60.06 )
};

%ETA
\addplot [dashed,mark options={scale=1,solid}, color=black, mark=o, mark size=1, mark options={scale=2,solid}] coordinates {
( -28.5 ,  -30 )
( -33.66 ,  -35.06 )
( -39.11 ,  -40.04 )
( -44.6 ,  -45.01 )
( -51.03 ,  -50.03 )
( - 58.6 ,  -55.0)
};

%MA1
\addplot [dashed,mark options={scale=1,solid},color=red, mark=diamond*, mark size=2.5] coordinates {
( -30.21  ,  -30.14 )
( -35.05 ,  -35.2073 )
( -40.33 ,  -40.3692 )
( -45.131 ,  -45.1124 )
( -50.18 ,  -50.0726 )
( -54.41  ,  -55.0344 )
( -58.27 ,  -60.0366 )
};
\legend{Trunc,AMA5,LOA,ETA-I,MA}
\end{axis}
\end{tikzpicture}
\subcaption{Parameterized error model}
  \label{fig:fir}
\end{subfigure}
\caption{Comparison of simulated noise power and analytically computed noise power for FIR filter when input static probability is (a) assumed to be 0.5 (b) computed as in Section~\ref{piAdder} for various approximate adders.}
  \label{fig:sim_anal_fir}
\end{figure}
\begin{figure}[!tb]
\begin{subfigure}{0.48\columnwidth}
\begin{tikzpicture}[scale=0.525]
\begin{axis}[
legend columns = 3,
every axis plot/.append style={ultra thick},
y label style={at={(0.05,0.5)}},
     xmin=-62, xmax=-28,
     ymin=-62, ymax=-28,
     xlabel= Actual noise power (dB),
     ylabel= Analytical noise power (dB),
label style={font=\large},
                    tick label style={font=\small} ,
    ymajorgrids=true,
    grid style=dashed,
	grid=major,
	legend style={font=\small,at={(0,1.0)},anchor=north west}] %($\mu$ W)]
%Trunc_pi05
\addplot [dashed,mark options={scale=1,solid},color=cyan, mark=square, mark size=1.5] coordinates {
( -31.12 , -30.0807 )
( -35.558 ,  -35.2364 )
( -41.145 ,  -40.5126 )
( -46.322 ,  -45.2087 )
( -52.469 ,  -50.3565 )
( -59.733 ,  -55.456 )
%( -71.114 ,  -60.6916 )
};

%AMA5
\addplot [dashed,mark options={scale=1,solid}, color=orange,mark=*, mark size=1, mark options={scale=2,solid}] coordinates {
( -30.867 ,  -30.3508 )
( -35.135 ,  -35.0352 )
( -40.434 ,  -40.0609 )
( -44.95 ,  -45.1321 )
( -50.284 ,  -50.1341 )
( -55.582 ,  -55.0257 )
( -60.187 ,  -60.0101 )
};

%LOA
\addplot [dashed,mark options={scale=1,solid}, color=teal, mark=star, mark size=1.75, mark options={scale=2,solid}] coordinates {
( -28.33,  -30.01 )
( -31.18 ,  -35.03 )
( -39.12 ,  -40.02 )
( -42.69 ,  -45.16 )
( -46.26 ,  -50.12 )
( -52.55 ,  -55.16 )
( -59.99 ,  -60.09 )
};

%ETA
\addplot [dashed,mark options={scale=1,solid}, color=black, mark=o, mark size=1, mark options={scale=2,solid}] coordinates {
( -28.98 ,  -30.02 )
( -33.51 ,  -35.09 )
( -38.68 ,  -40.1 )
( -44.07 ,  -45.13 )
( -50.49 ,  -50.1 )
( -59.25 ,  -55.0 )
%( -78.6 ,  -60.44 )
};

%MA1
\addplot [dashed,mark options={scale=1,solid},color=red, mark=diamond*, mark size=2.5] coordinates {
%( -20.4 ,  -30.12 )
%( -26.88 , -35.03 )
( -29.37 , -40.06 )
( -34.88 , -45.34 )
( -47.15 , -50.0 )
( -49.07 , -55 )
( -54.89 , -60.1 )
};

\legend{Trunc,AMA5,LOA,ETA-I,MA}
\end{axis}
\end{tikzpicture}
\subcaption{Uniform distribution}
  \label{fig:WL_fir}
\end{subfigure}
\begin{subfigure}{0.48\columnwidth}
  \centering
\begin{tikzpicture}[scale=0.525]
   \begin{axis}[
legend columns = 3,
every axis plot/.append style={ultra thick},
y label style={at={(0.07,0.5)}},
     xmin=-62, xmax=-28,
     ymin=-62, ymax=-28,
     xlabel= Actual noise power (dB),
     ylabel= Analytical noise power (dB),
label style={font=\large},
                    tick label style={font=\small} ,
    ymajorgrids=true,
    grid style=dashed,
	grid=major,
	legend style={font=\small,at={(1,0)},anchor=south east}] %($\mu$ W)]
%Trunc
\addplot [dashed,mark options={scale=1,solid},color=cyan, mark=square, mark size=1.5] coordinates {
( -31.12 , -30.0807 )
( -35.558 ,  -35.2364 )
( -41.145 ,  -40.5126 )
( -46.322 ,  -45.2087 )
( -52.469 ,  -50.3565 )
( -59.733 ,  -55.456 )
%( -71.114 ,  -60.6916 )
};

%AMA5
\addplot [dashed,mark options={scale=1,solid}, color=orange,mark=*, mark size=1, mark options={scale=2,solid}] coordinates {
( -30.867 ,  -30.3508 )
( -35.135 ,  -35.0352 )
( -40.434 ,  -40.0609 )
( -44.95 ,  -45.1321 )
( -50.284 ,  -50.1341 )
( -55.582 ,  -55.0257 )
( -60.187 ,  -60.0101 )
};
%LOA
\addplot [dashed,mark options={scale=1,solid}, color=teal, mark=star, mark size=1.75, mark options={scale=2,solid}] coordinates {
( -29.664 ,  -30.1428 )
( -35.219 ,  -35.0315 )
( -39.531 ,  -40.1945 )
( -44.824 ,  -45.1407 )
( -50.645 ,  -50.6035 )
( -54.37 ,  -55.0699 )
( -59.564 ,  -60.0466 )
};
%ETA
\addplot [dashed,mark options={scale=1,solid}, color=black, mark=o, mark size=1, mark options={scale=2,solid}] coordinates {
( -31.319 ,  -30.0693 )
( -34.033 ,  -35.1208 )
( -40.327 ,  -40.0183 )
( -45.824 ,  -45.0477 )
( -52.498 ,  -50.0109 )
( -55.1 ,  -55.0309 )
( -60.165 ,  -60.3507 )
};

%Median
\addplot [dashed,mark options={scale=1,solid},color=red, mark=diamond*, mark size=2.5] coordinates {
( -30.243 ,  -30.4476 )
( -35.547 ,  -35.2317 )
( -40.316 ,  -40.0734 )
( -45.215 ,  -46.0667 )
( -49.796 ,  -50.0031 )
( -55.393 ,  -55.2387 )
( -57.93  ,  -60.1156 )
};
\legend{Trunc,AMA5,LOA,ETA-I,MA}
\end{axis}
\end{tikzpicture}
  \subcaption{Parameterized error model}
  \label{fig:iir}
\end{subfigure}
\caption{Comparison of simulated noise power and analytically computed noise power for IIR filter when input static probability is (a) assumed to be 0.5 (b) computed as in Section~\ref{piAdder} for various approximate adders.}
  \label{fig:sim_anal_iir}
\end{figure}

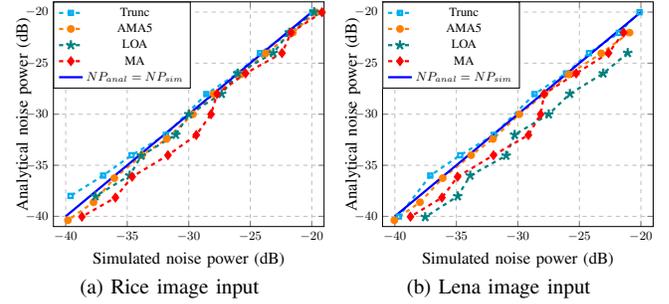
\begin{figure}[!tb]
\begin{subfigure}{0.48\columnwidth}
  \centering
\begin{tikzpicture}[scale=0.525]
   \begin{axis}[
legend columns = 1,
every axis plot/.append style={ultra thick},
y label style={at={(0.07,0.5)}},
     xmin=-41, xmax=-19,
     ymin=-41, ymax=-19,
     xlabel= Simulated noise power (dB),
     ylabel= Analytical noise power (dB),
label style={font=\large},
                    tick label style={font=\small} ,
    ymajorgrids=true,
    grid style=dashed,
	grid=major,
	legend style={font=\small,at={(0,1.0)},anchor=north west}] %($\mu$ W)]

%Trunc
\addplot [dashed,mark options={scale=1,solid},color=cyan, mark=square, mark size=1.5] coordinates {
( -19.96 ,  -20.0111 )
( -21.81 ,  -22.0347 )
( -24.21 ,  -24.02 )
( -26.03,  -26.0043 )
( -28.57 ,  -28.0258 )
%( -27.31 ,  -30.021 )
( -31.86 ,  -32.024 )
( -34.65, -34.0182 )
( -36.96, -36.0067 )
( -39.61, -38.0002 )
%( -39.67, -40.0187 )
};
%AMA5
\addplot [dashed,mark options={scale=1,solid}, color=orange,mark=*, mark size=1, mark options={scale=2,solid}] coordinates {
( -19.67 ,  -20.0223 )
( -21.54 ,  -22.0106 )
( -23.74 ,  -24.0402 )
( -25.75 ,  -26.0965 )
( -27.98 ,  -28.0154 )
( -29.66 ,  -30.0024)
( -31.79 ,  -32.4245 )
( -33.79 ,  -34.0151)
( -36.05 ,  -36.2615)
( -37.74 ,  -38.6224)
( -39.82 ,  -40.3833)
};
%LOA
\addplot [dashed,mark options={scale=1,solid}, color=teal, mark=star, mark size=1.75, mark options={scale=2,solid}] coordinates {
( -19.804 ,  -20.0015 )
( -21.825 ,  -22.005 )
( -23.127 ,  -24.0004  )
( -26.027 ,  -26.0028 )
( -27.327 ,  -28.0043 )
( -29.989 ,  -30.0043 )
( -31.05 ,  -32.0075 )
( -33.851 ,  -34.0557)
( -34.8 ,  -36.0015)
( -37.59 ,  -38.0303)
};
%Median
\addplot [dashed,mark options={scale=1,solid},color=red, mark=diamond*, mark size=2.5] coordinates {
( -19.17 ,  -20.019 )
( -21.68 ,  -22.0053 )
( -22.42 ,  -24.0151 )
( -25.38 ,  -26.0031 )
( -27.7 ,  -28.0154 )
( -28.19 ,  -30.0024 )
( -29.39 ,  -32.037 )
( -31.69 ,  -34.0151 ) 
( -34.6 ,  -36.1236 )
( -35.96 ,  -38.1648 )
( -38.685 ,  -40.0357 )
};
%45 degree line
\addplot [mark options={scale=1,solid},color=blue,  mark size=1.5] coordinates {
( -20,-20)
( -22,-22 )
( -24,-24 )
( -26,-26 )
( -28,-28 )
( -30,-30 )
( -32,-32 )
( -34,-34 )
( -36,-36 )
( -38,-38 )
( -40,-40 )
};
\legend{Trunc,AMA5,LOA,MA,$NP_{anal}=NP_{sim}$}
\end{axis}
\end{tikzpicture}
  \subcaption{Rice image input}
  \label{fig:dct_Rice}
\end{subfigure}
\begin{subfigure}{0.48\columnwidth}
  \centering
\begin{tikzpicture}[scale=0.525]
   \begin{axis}[
legend columns = 1,
every axis plot/.append style={ultra thick},
y label style={at={(0.07,0.5)}},
     xmin=-41, xmax=-19,
     ymin=-41, ymax=-19,
     xlabel= Simulated noise power (dB),
     ylabel= Analytical noise power (dB),
label style={font=\large},
                    tick label style={font=\small} ,
    ymajorgrids=true,
    grid style=dashed,
	grid=major,
	legend style={font=\small,at={(0,1.0)},anchor=north west}] %($\mu$ W)]

%Trunc
\addplot [dashed,mark options={scale=1,solid},color=cyan, mark=square, mark size=1.5] coordinates {
( -20.11 ,  -20.0111 )
( -21.79 ,  -22.0347 )
( -24.22 ,  -24.02 )
( -26.09,  -26.0043 )
( -28.64 ,  -28.0258 )
%( -27.29 ,  -30.021 )
( -31.96 ,  -32.024 )
( -34.69, -34.0182 )
( -37.12, -36.0067 )
%( -39.55, -38.0002 )
( -39.67, -40.0187 )
};
%AMA5
\addplot [dashed,mark options={scale=1,solid}, color=orange,mark=*, mark size=1, mark options={scale=2,solid}] coordinates {
%( -19.21 ,  -20.0223 )
( -20.92 ,  -22.0106 )
( -23.22 ,  -24.0402 )
( -25.86 ,  -26.0965 )
( -27.83 ,  -28.0154 )
( -29.88 ,  -30.0024)
( -32.02 ,  -32.4245 )
( -33.8 ,  -34.0151)
( -36.11 ,  -36.2615)
( -37.96 ,  -38.6224)
( -40.04 ,  -40.3833)
};
%LOA
\addplot [dashed,mark options={scale=1,solid}, color=teal, mark=star, mark size=1.75, mark options={scale=2,solid}] coordinates {
%( -15.94 ,  -20.0015 )
%( -19.53 ,  -22.005 )
( -21.07 ,  -24.0004  )
( -23.06 ,  -26.0028 )
( -25.77 ,  -28.0043 )
( -27.49 ,  -30.0043 )
( -30.26 ,  -32.0075 )
( -30.95 ,  -34.0557)
( -33.87 ,  -36.0015)
( -34.89 ,  -38.0303)
( -37.53 ,  -40.0669)
};

%Median
\addplot [dashed,mark options={scale=1,solid},color=red, mark=diamond*, mark size=2.5] coordinates {
%( -18.29 ,  -20.019 )
( -21.4 ,  -22.0053 )
( -22.63 ,  -24.0151 )
( -25.26 ,  -26.0031 )
( -27.86 ,  -28.0154 )
( -28.15 ,  -30.0024 )
( -29.14 ,  -32.037 )
( -31.98 ,  -34.0151 ) 
( -34.9 ,  -36.1236 )
( -36.17 ,  -38.1648 )
( -38.73 ,  -40.0357 )
};
%45 degree line
\addplot [mark options={scale=1,solid},color=blue,  mark size=1.5] coordinates {
( -20,-20)
( -22,-22 )
( -24,-24 )
( -26,-26 )
( -28,-28 )
( -30,-30 )
( -32,-32 )
( -34,-34 )
( -36,-36 )
( -38,-38 )
( -40,-40 )
};
\legend{Trunc,AMA5,LOA,MA,$NP_{anal}=NP_{sim}$}
\end{axis}
\end{tikzpicture}
  \subcaption{Lena image input}
  \label{fig:dct_Lena}
\end{subfigure}
\caption{Comparison of simulated noise power and analytically computed noise power for $8 \times 8$ DCT using parameterized error model.}
  \label{fig:sim_anal_dct}
\end{figure}

The DCT module \cite{BAS09} is a multiplierless implementation with adders and subtractors. 
Fig.~\ref{fig:dct_Rice} and Fig.~\ref{fig:dct_Lena} 
show the comparison of simulated value of noise power %obtained using Rice image and Lena image as inputs 
and analytically computed noise power using the parameterized error model for various approximate adders, for two different images namely Rice image and Lena image respectively. It can be seen that the analytical values match reasonably well with the actual values (with a maximum of 2.5\,dB and 3\,dB error for Rice and Lena images respectively).
%This is because the inputs are image pixel intensities and upto 6 bits are approximated in some cases for high values of noise power. %This error reduces if the primary inputs are uniformly distributed.

%%%%%%%%%%%%%%%%%%%%%%%%%%%%%%%%%%%%%%%%%%%%%%%%%%%%%%%%%%%%%%%%%%%%%%%%%%%%%%%%%%%%%%%%%%%%%%%%%%%%%%%%%%

\section{Conclusion}
\label{conclusion}
We have proposed parameterized error models for approximate adders using input static probabilities as
parameters and incorporated these error models in an optimization framework. We have shown that the parameterized error models
provide better noise power prediction than the typical error
models that assume uniform input distribution. The results of FIR and IIR filters and DCT module show that the use of parameterized error model in the optimization framework improves the accurate prediction of the overall MSE.% We use this in an optimization framework and show that the  optimizer requires the parameterized error models with the correct values of the static probability for accurate prediction of the MSE.

\bibliographystyle{ieeetr}
\bibliography{ref}

\begin{thebibliography}{10}

\bibitem{GeAr2015}
M.~Shafique, W.~Ahmad, R.~Hafiz, and J.~Henkel, ``A low latency generic
  accuracy configurable adder,'' in {\em Proceedings of the 52Nd Annual Design
  Automation Conference}, DAC '15, (New York, NY, USA), ACM, 2015.

\bibitem{inexactadders2016}
H.~A.~F. Almurib, T.~N. Kumar, and F.~Lombardi, ``Inexact designs for
  approximate low power addition by cell replacement,'' in {\em Design,
  Automation and Test in Europe (DATE)}, 2016.

\bibitem{XORbased2013}
Z.~Yang, A.~Jain, J.~Liang, J.~Han, and F.~Lombardi, ``Approximate
  xor/xnor-based adders for inexact computing,'' {\em 2013 13th IEEE
  International Conference on Nanotechnology (IEEE-NANO 2013)}, 2013.

\bibitem{AMA2013}
V.~Gupta, D.~Mohapatra, A.~Raghunathan, and K.~Roy, ``Low-power digital signal
  processing using approximate adders,'' {\em IEEE Trans. on Comp.-Aided Design
  of Integrated Circuits and Systems}, vol.~32, 1 2013.

\bibitem{LOA2010}
H.~R. Mahdiani, A.~Ahmadi, S.~M. Fakhraie, and C.~Lucas, ``Bio-inspired
  {Imprecise} computational blocks for efficient {VLSI} implementation of
  soft-computing applications,'' {\em IEEE Trans. on Circuits and Systems I:
  Regular Papers}, vol.~57, 4 2010.

\bibitem{zhu2010}
N.~Zhu, W.~L. Goh, W.~Zhang, K.~S. Yeo, and Z.~H. Kong, ``Design of low-power
  high-speed truncation-error-tolerant adder and its application in digital
  signal processing,'' {\em IEEE TVLSI}, vol.~18, no.~8, 2010.

\bibitem{CeliaDATE2018}
D.~Celia, V.~Vasudevan, and N.~Chandrachoodan, ``Optimizing power-accuracy
  trade-off in approximate adders,'' in {\em 2018 Design, Automation Test in
  Europe Conference Exhibition (DATE)}, March 2018.

\bibitem{FIR_newCAS2015}
L.~B. {Soares}, S.~{Bampi}, and E.~{Costa}, ``Approximate adder synthesis for
  area- and energy-efficient fir filters in cmos vlsi,'' in {\em 2015 IEEE 13th
  International New Circuits and Systems Conference}, June 2015.

\bibitem{JPEGDAC2016}
F.~S. Snigdha, D.~Sengupta, J.~Hu, and S.~S. Sapatnekar, ``Optimal design of
  jpeg hardware under the approximate computing paradigm,'' in {\em 2016 53nd
  ACM/EDAC/IEEE DAC}, June 2016.

\bibitem{DCT_DATE2017}
Z.~{Vasicek}, V.~{Mrazek}, and L.~S. {Brno}, ``Towards low power approximate
  dct architecture for hevc standard,'' in {\em Design, Automation Test in
  Europe Conference Exhibition (DATE), 2017}, March 2017.

\bibitem{SABER_DAC2017}
D.~{Sengupta}, F.~S. {Snigdha}, {Jiang Hu}, and S.~S. {Sapatnekar}, ``Saber:
  Selection of approximate bits for the design of error tolerant circuits,'' in
  {\em 2017 54th ACM/EDAC/IEEE DAC}, June 2017.

\bibitem{Framework_TCASI2019}
M.~{Pashaeifar}, M.~{Kamal}, A.~{Afzali-Kusha}, and M.~{Pedram}, ``A
  theoretical framework for quality estimation and optimization of dsp
  applications using low-power approximate adders,'' {\em IEEE Trans. on
  Circuits and Systems I: Regular Papers}, vol.~66, Jan 2019.

\bibitem{TGA2015}
Z.~{Yang}, J.~{Han}, and F.~{Lombardi}, ``Transmission gate-based approximate
  adders for inexact computing,'' in {\em Proceedings of the 2015 IEEE/ACM
  NANOARCH´15}, July 2015.

\bibitem{BAS09}
S.~{Bouguezel}, M.~O. {Ahmad}, and M.~N.~S. {Swamy}, ``A fast $8\times 8$
  transform for image compression,'' in {\em 2009 International Conference on
  Microelectronics - ICM}, Dec 2009.

\bibitem{ErrorPMF_TCAD2019}
D.~{Sengupta}, F.~S. {Snigdha}, J.~{Hu}, and S.~S. {Sapatnekar}, ``An
  analytical approach for error pmf characterization in approximate circuits,''
  {\em IEEE Trans. on CAD of Integrated Circuits and Systems}, vol.~38, Jan
  2019.

\bibitem{CeliaISCAS2018}
D.~Celia, V.~Vasudevan, and N.~Chandrachoodan, ``Probabilistic error modeling
  for two-part segmented approximate adders,'' in {\em 2018 IEEE International
  Symposium on Circuits and Systems (ISCAS)}, May 2018.

\bibitem{Qnoise1977}
A.~{Sripad} and D.~{Snyder}, ``A necessary and sufficient condition for
  quantization errors to be uniform and white,'' {\em IEEE Trans. on Acoustics,
  Speech, and Signal Processing}, vol.~25, October 1977.

\bibitem{WLO}
D.~Menard, N.~Herve, O.~Sentieys, and H.-N. Nguyen, ``High-level synthesis
  under fixed-point accuracy constraint,'' {\em Journal of Electrical and
  Computer Engineering}, vol.~2012, Jan. 2012.

\end{thebibliography}

%\onecolumn
%\input{tcas2_resp.tex}

\end{document}